\documentclass[aps,prb,twocolumn,floatfix,amsmath,amssymb,showpacs,
               superscriptaddress,10pt]{revtex4-1}
\usepackage[final]{graphicx}
\usepackage{placeins} 
\usepackage{float}
\usepackage{color}
\usepackage{dcolumn}
\usepackage{xcolor}
\usepackage{url}
\graphicspath{{Figures/}}
\usepackage{bm}
\usepackage{epsfig,psfrag,amsmath,amssymb}
\input{epsf}
\usepackage{hyperref}
\usepackage[percent]{overpic}

\hypersetup{
    colorlinks=true,
    linkcolor=blue,
    citecolor=blue,
    filecolor=magenta,      
    urlcolor=blue,
}
\usepackage{array}
\usepackage{tabu}
\newcolumntype{C}[1]{>{\centering\arraybackslash$}p{#1}<{$}}

\setcitestyle{square}

\bibpunct{[}{]}{,}{n}{}{}
 
\begin{document}
\title{Multitude of Topological Phase Transitions in Bipartite \\ Dice and Lieb  Lattices with Interacting Electrons and Rashba Coupling}

\author{Rahul Soni}
\affiliation{Department of Physics and Astronomy, The University of 
Tennessee, Knoxville, Tennessee 37996, USA}
\affiliation{Materials Science and Technology Division, Oak Ridge National 
Laboratory, Oak Ridge, Tennessee 37831, USA}
\author{Amit Bikram Sanyal}
\affiliation{Department of Physics and Astronomy, The University of 
Tennessee, Knoxville, Tennessee 37996, USA}
\author{Nitin Kaushal}
\affiliation{Materials Science and Technology Division, Oak Ridge National 
Laboratory, Oak Ridge, Tennessee 37831, USA}
\author{Satoshi Okamoto}
\affiliation{Materials Science and Technology Division, Oak Ridge National 
Laboratory, Oak Ridge, Tennessee 37831, USA}
\author{Adriana Moreo}
\affiliation{Department of Physics and Astronomy, The University of 
Tennessee, Knoxville, Tennessee 37996, USA}
\affiliation{Materials Science and Technology Division, Oak Ridge National 
Laboratory, Oak Ridge, Tennessee 37831, USA}
\author{Elbio Dagotto}
\affiliation{Department of Physics and Astronomy, The University of 
Tennessee, Knoxville, Tennessee 37996, USA}
\affiliation{Materials Science and Technology Division, Oak Ridge National 
Laboratory, Oak Ridge, Tennessee 37831, USA}
\date{\today}

\begin{abstract}
We report the results of a Hartree-Fock study applied to interacting electrons moving in two different bipartite lattices: the dice and the Lieb lattices, at half-filling. Both lattices develop ferrimagnetic order in the phase diagram $U$-$\lambda$, where $U$ is the Hubbard onsite repulsion and $\lambda$ the Rashba spin-orbit coupling strength. Our main result is the observation of 
an unexpected multitude of topological phases for both lattices. All these phases are ferrimagnetic, but they differ among themselves 
in their set of six Chern numbers (six numbers because the
unit cells have three atoms). The Chern numbers $|C|$ observed in our study range from 0 to 3, showing that large Chern numbers can be obtained by the effect of electronic correlations, adding to the recently discussed methodologies to increase $|C|$ based on extending the hopping range in tight-binding models, using sudden quenches, or photonic crystals, all 
without including electronic interactions.   
\end{abstract}
\maketitle

\section{Introduction}

In the seminal paper of Haldane, a tight-binding model on a honeycomb lattice with a staggered flux
was shown to induce the integer quantum Hall effect~\cite{haldane88}, 
without the need for external magnetic fields.
Subsequent generalizations led to the concept of topological 
insulators~\cite{top-ins1,top-ins2,top-ins3}, widely studied at present. 
A commonly used methodology employed in this area of research is to search for models that display
quasi-two-dimensional flat bands with a nonzero Chern number, in the presence of external magnetic fields. This Chern number equals the number $|C|$ of chiral modes at the edges if open boundary conditions are used. The bulk is insulating, while the edge is conducting via those
chiral modes. The sign of $C$ reflects on the sense of edge mode circulation, clockwise or anticlockwise. 
There are other symmetry protected insulators with robust edge states, such as in quantum spin Hall insulators~\cite{QSHI} and topological superconductors with Majorana fermions at the edges~\cite{1D-Majorana}, all promising candidates for quantum computation because the symmetry-protected edge states are not affected from back-scattering.

A generalization of the honeycomb model of Haldane leads to the {\it dice} lattice via the addition of an extra
site at the center of each hexagon. This lattice has an entire flat band of localized states~\cite{sutherland,wang11} 
(see also Refs.\cite{dice-previous1,dice-previous2,dice-previous3}). As early as 
the 90's and early 00's the dice lattice was studied in the context of Josephson Junction 
Arrays and bosonic systems, already predicting three flat
bands in a magnetic field~\cite{vidal}, subsequently confirmed via transport measurements 
using superconducting wire networks~\cite{abilio}. The {\it Lieb} 
lattice~\cite{lieb-original} is also receiving renewed attention due to its flat
band and potential connection to superconductivity via the novel concept of quantum geometry~\cite{bernevig}.

\begin{figure}[!t] 
\centering
\includegraphics[width=3.1in, height=4.0in]{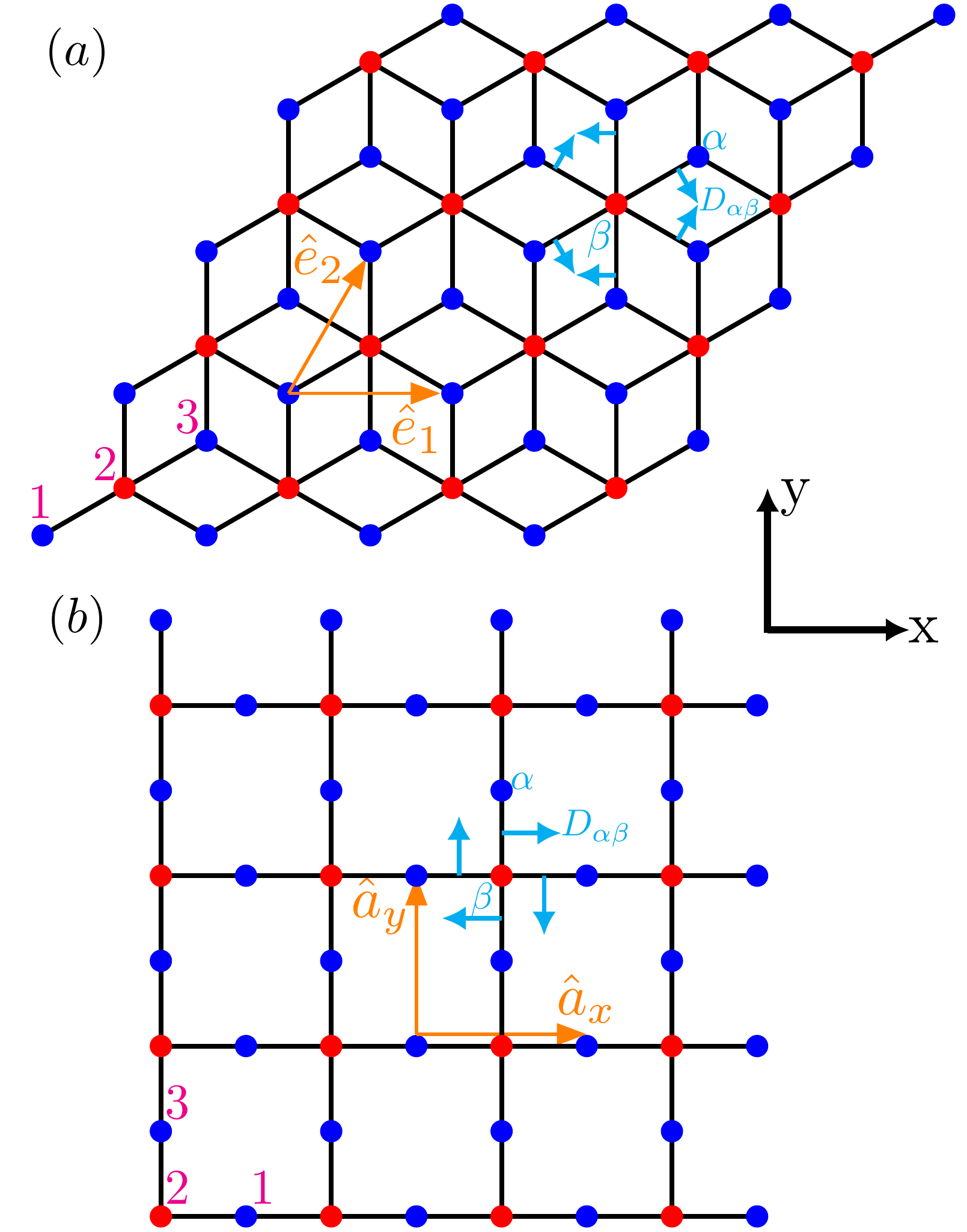}
\caption{(a) Geometry of the dice lattice. Here we show a 4$\times$4 cluster with 4 unit cells along each lattice unit vectors $\hat{e}_1$ and $\hat{e}_2$ (in orange). The blue (red) sites have coordination 3 (6). (b) Geometry of a 4$\times$4 Lieb cluster with 4 unit cells along each lattice vectors $\hat{a}_x$ and $\hat{a}_y$ (in orange), respectively. The blue (red) sites have coordination 2 (4). Each unit cell contains 3 sites, marked as $1,2,3$. In both (a) and (b), the cyan arrows labelled $\hat{D}_{\alpha\beta}$ indicate the Rashba SOC directions on bonds $\alpha\beta$.\label{Fig: Geometries}}
\end{figure}

The dice lattice has two
types of sites: some with coordination three and others
with coordination six, as shown in Fig.~\ref{Fig: Geometries}(a). The unit cell contains three sites, 
leading to three bands, each one doubly degenerate due to spin. 
The noninteracting tight-binding model on a dice lattice including Rashba spin-orbit coupling
and in the presence of magnetic fields to break the degeneracy (thus having a total of six bands) leads to a half-filled ground state with $|C|=2$~\cite{wang11}, larger than the $|C|=1$ of the Haldane model 
(see also Ref.\cite{cook14}). Physical realizations of this lattice are possible. For example,
bulk oxides with the generic formula A$_4$B$^{'}$B$_2$O$_{12}$, such as Ba$_4$CoRe$_2$O$_{12}$~\cite{zhou}, contain trilayers that seen from above resemble a dice lattice.

In a previous publication by our group we studied ribbons of dice lattice~\cite{soni20}, 
equivalent to a dimensional reduction from two to one of the original dice lattice into a 
quasi-one-dimensional system. Qualitatively, ribbons were shown to behave very
similarly to planar dice lattices~\cite{soni20}.
This paves the way towards the introduction of electronic correlations,
which is a much simpler task in one-dimensional systems than in planes due to the availability of
many-body techniques that are particularly efficient in one dimension.  
However, carrying out density matrix renormalization group (DMRG)~\cite{DMRGreview} 
or Lanczos diagonalization~\cite{lanczos} studies of interacting electrons in dice ribbons is still a 
considerable computational challenge. For this reason, in this publication, as an intermediate step towards the
full introduction of correlations and quantum fluctuations, we employ
the self-consistent Hartree Fock approximation to directly study dice planes instead of ribbons.

In the present effort, we also study the Lieb lattice, 
Fig.~\ref{Fig: Geometries}(b). Besides being bipartite like the dice lattice, we will show it shares many
similar properties in the phase diagram with the dice lattice. Lieb lattices have been
realized in optical lattices~\cite{lieb-optical1,lieb-optical2} and photonic crystals~\cite{lieb-photonic1,lieb-photonic2}. The study of Lieb lattices including intrinsic SOC and $U$ has been discussed in \cite{jana19}.

Most early theoretical work in this context have reported Chern numbers $|C|$ equal to 0, 1, or 2 in absolute value.
However, having even larger Chern numbers can provide practical improvements in potential applications. Because of the 
correspondence between $|C|$ and the number of dissipationless edge modes, the performance of devices can be improved
by reducing the contact resistance in the quantum anomalous Hall effect. Recent efforts have found procedures 
to increase the Chern number. For example, (1) by considering hoppings at longer distances than those already 
contained in the original tight-binding Haldane model with $|C|=1$, a multiplication of Dirac points can be  
achieved~\cite{bena11}. A general procedure
to construct Chern insulators with arbitrary $|C|=n$, with $n$ an integer, employing extended hopping
interactions was presented in Ref.~\cite{sticlet12}. Chern numbers as large as $|C|=5$ were reported~\cite{sticlet17}. 
(2) In the context of photonic crystals using multimode one-way waveguides, 
$|C|$ as large as 4 has been reached using {\it ab initio} calculations~\cite{skirloPRL}.
(3) Employing periodic quenching, and a two-band model as example, it was
shown that Chern numbers as large as 7 can be obtained~\cite{xiong2016}. 
(4) Sudden quenches can also modify Chern numbers. For example a system with $C=2$, with
two edge states, after a sudden quench to the nontopological regime with $C=0$, can have an intermediate phase
with $C=1$ due to different decay rates of the inner and outer edge modes~\cite{Bhatta}. In partial summary,
having a large $C$ is potentially beneficial for applications, and procedures to reach such goal have been
recently proposed, as the partial list provided above shows.

All these previous efforts have neglected electronic-electronic correlations primarily because already a rich
variety of topological phases can be obtained at the level of non-interacting electrons. Moreover, these models
can be solved exactly. However, the neglect of electron-electron interactions is always an approximation~\cite{rachel}. In addition,
correlation effects may induce novel phases, difficult to anticipate from the noninteracting limit. Consequently,
it is widely believed that the next big
challenge in quantum materials is the mixture of topological and correlation effects. Will they compete or cooperate?
What new phases will emerge adding correlation effects?
The main technical difficulty in this context is that electronic correlations substantially increase the difficulty in solving the Hamiltonian that now contains both Hubbard $U$ interactions and spin-orbit coupling $\lambda$.

In this publication, we study the dice and Lieb lattices in the presence of onsite Hubbard $U$ repulsion, within the Hartree Fock (HF) approximation. We present the results for these two lattices in the same publication because of their
many similarities: both have unit cells with 3 sites (2 of those sites equivalent by symmetry) and both develop nonzero Chern numbers in the noninteracting $U=0$ limit when in the presence of spin-orbit Rashba interactions of coupling 
strength $\lambda$ and an external magnetic field. By solving the self-consistent equations numerically we find two main results: (1) Both lattices develop {\it ferrimagnetic} order. This confirms previous studies by our group carried out by Lanczos on 2$\times$2 unit cells (i.e. 12 sites) and DMRG
on 2$\times$8 ribbons (48 sites), where ferrimagnetic order was found. Moreover, the ferrimagnetic order develops
immediately turning on $U$, in qualitative agreement also with the small cluster studies mentioned above. The flat bands in the non-interacting limit~\cite{sutherland}, without external fields, split in the presence of the Hubbard interaction. This HF analysis confirms the previous conjecture~\cite{soni20} that using ribbons to study properties of planes is qualitatively, and often quantitatively, correct. (2) More importantly, we here report unexpectedly rich phase diagrams varying $U$ and $\lambda$, unveiling a plethora of phases with a variety of Chern numbers, some as large as $|C|$=3. Thus, not only by increasing the range of hoppings or by quenched-dynamic setups is that $|C|$ can be increased, but our results suggest the presence of strong correlation $U$ can lead to similar effects, at least within the HF approximation.

The organization of the paper is as follows. In Sec.~II, the Model and Method are described, including the Fourier transform of the Hamiltonian which amounts to a 6$\times$6 matrix at each fixed momentum. In this section,
the HF approximation is also explained, as well as the technique to iteratively find the order parameters self-consistently.
In Sec.~III the results are discussed, separated into dice and Lieb lattices subsections, both containing phase
diagrams with the many topological phases we found. Finally, Sec.~IV contains the Conclusions.

\section{Model and Method}
\subsection{Non-Interacting Electrons with Spin-Orbit Coupling}
The non-interacting Hamiltonian comprises of the tight-binding kinetic energy term and the Rashba spin-orbit coupling term. These Hamiltonians have been studied before in Refs.~\cite{wang11,soni20} for the dice lattice and in Ref.~\cite{chen17} for the Lieb lattice. The non-interacting Hamiltonian for the dice and Lieb lattices is defined as
\begin{eqnarray}
H_{Dice(Lieb)} &=& -t\sum_{\substack{\mathbf{r},\mathbf{r}',\\ \alpha,\beta,\sigma}}\left(c^{\dagger}_{\mathbf{r},\alpha,\sigma}c^{\phantom{dagger}}_{\mathbf{r}',\beta,\sigma}+h.c.\right) -\epsilon\sum_{\mathbf{r}} n_{\mathbf{r},2} \nonumber \\
& &-\lambda\sum_{\substack{\mathbf{r},\mathbf{r}',\\ \alpha,\beta,\sigma,\sigma'}}\left(\mathfrak{i} c^{\dagger}_{\mathbf{r},\alpha,\sigma}(\hat{D}_{\alpha\beta}.\vec{\tau})_{\sigma\sigma'}c^{\phantom{dagger}}_{\mathbf{r}',\beta,\sigma'}+h.c.\right) \nonumber \\ \label{Eqn: NI Ham Real Space}
\end{eqnarray}

where $\mathbf{r}, \mathbf{r}'$ are the unit cell indexes, $\alpha$ and $\beta$ are the site indexes within the unit cell $\mathbf{r}$ and $\mathbf{r}'$, respectively (with $\alpha,\beta=1,2,3$), and $\sigma=\uparrow, \downarrow$ is the $z$-axis spin projection of the electron at site $\alpha$ within the unit cell $\mathbf{r}$. $\lambda$ is the Rashba spin-orbit coupling strength that is  uniform for all the bonds, while $\epsilon$ is the onsite energy that affects only the red sites of Fig.~\ref{Fig: Geometries}. $\vec{\tau}=\tau_{x}\hat{x} + \tau_{y}\hat{y} + \tau_{z}\hat{z}$ is the Pauli matrix vector, $\hat{D}_{\alpha\beta}$ is the unit vector in-plane and perpendicular to the bond formed by $(\mathbf{r},\alpha)$ and $(\mathbf{r}',\beta)$. Both Rashba and hopping occur only between nearest-neighbor sites. Note that the $\hat{D}_{\alpha\beta}$ for the dice lattice in Fig.~\ref{Fig: Geometries} follows the $D_{3d}$ symmetry group~\cite{wang11}. \\

\noindent
Via the Fourier transform $c^{\dagger}_{\mathbf{k},\alpha,\sigma}=\frac{1}{\sqrt{N_1N_2}}\sum_{\mathbf{r}}e^{\mathfrak{i}\mathbf{k}.\mathbf{r}}c^{\dagger}_{\mathbf{r},\alpha,\sigma}$ the non-interacting Hamiltonian of the dice lattice in momentum space \cite{wang11,soni20} becomes:
\begin{eqnarray}
&& \hspace*{2cm} H_{Dice}(\mathbf{k}) = \nonumber \\
&& \hspace*{-1cm}\begin{pmatrix}
0 & 0 & -t\gamma_{\mathbf{k}}^{*} & -\mathfrak{i}\lambda\gamma_{\mathbf{k}+}^{*} & 0 & 0\\
0 & 0 & -\mathfrak{i}\lambda\gamma_{\mathbf{k}-}^{*} & -t\gamma_{\mathbf{k}}^{*} & 0 & 0\\
-t\gamma_{\mathbf{k}} & \mathfrak{i}\lambda\gamma_{\mathbf{k}-} & -\epsilon & 0 & -t\gamma_{\mathbf{k}}^{*} & \mathfrak{i}\lambda\gamma_{\mathbf{k}+}^{*} \\
\mathfrak{i}\lambda\gamma_{\mathbf{k}+} & -t\gamma_{\mathbf{k}} & 0 & -\epsilon & \mathfrak{i}\lambda\gamma_{\mathbf{k}-}^{*} & -t\gamma_{\mathbf{k}}^{*}  \\
0 & 0 & -t\gamma_{\mathbf{k}} & -\mathfrak{i}\lambda\gamma_{\mathbf{k}-} & 0 & 0 \\
0 & 0 & -\mathfrak{i}\lambda\gamma_{\mathbf{k}+} & -t\gamma_{\mathbf{k}} & 0 & 0 \label{Eqn: Dice Ham K-Space}
\end{pmatrix} 
\end{eqnarray}

\noindent 
where $\gamma_{\mathbf{k}}$=$1+e^{\mathfrak{i}k_1}+e^{\mathfrak{i}k_2}$, and $\gamma_{\mathbf{k} \pm}$=$ 1+e^{\mathfrak{i}(k_1 \pm 2\pi/3)}+e^{\mathfrak{i}(k_2 \pm 4\pi/3)}$, with $k_i=\mathbf{k}.\hat{e}_{i}$. These two 
components are along the lattice vectors $\hat{e}_{1}$ and $\hat{e}_{2}$. The annihilation operator basis used here is $(c^{\phantom{dagger}}_{\mathbf{k},1,\uparrow}, \hspace{0.15cm} c^{\phantom{dagger}}_{\mathbf{k},1,\downarrow}, \hspace{0.15cm} c^{\phantom{dagger}}_{\mathbf{k},2,\uparrow}, \hspace{0.15cm} c^{\phantom{dagger}}_{\mathbf{k},2,\downarrow}, \hspace{0.15cm} c^{\phantom{dagger}}_{\mathbf{k},3,\uparrow}, \hspace{0.15cm} c^{\phantom{dagger}}_{\mathbf{k},3,\downarrow})$. $N_1$ and $N_2$ are the number of unit cells along the lattice vectors $\hat{e}_{1}$ and $\hat{e}_{2}$, respectively.

Similarly, the non-interacting Hamiltonian of the Lieb lattice, under the Fourier transform $c^{\dagger}_{\mathbf{k},\alpha,\sigma}=\frac{1}{\sqrt{N_xN_y}}\sum_{\mathbf{r}}e^{\mathfrak{i}\mathbf{k}.\mathbf{r}}c^{\dagger}_{\mathbf{r},\alpha,\sigma}$ \cite{chen17} becomes:
\begin{eqnarray}
&& \hspace*{2cm} H_{Lieb}(\mathbf{k}) = \\
&&\begin{pmatrix}
0 & 0 & -t\delta_{k_x+}^{*} & \lambda\delta_{k_x-}^{*} & 0 & 0\\
0 & 0 & -\lambda\delta_{k_x-}^{*} & -t\delta_{k_x+}^{*} & 0 & 0\\
-t\delta_{k_x+} & -\lambda\delta_{k_x-} & -\epsilon & 0 & -t\delta_{k_y+} & \mathfrak{i}\lambda\delta_{k_y-} \\
\lambda\delta_{k_x-} & -t\delta_{k_x+} & 0 & -\epsilon & \mathfrak{i}\lambda\delta_{k_y-} & -t\delta_{k_y+} \\
0 & 0 & -t\delta_{k_y+}^{*} & -\mathfrak{i}\lambda\delta_{k_y-}^{*} & 0 & 0 \\
0 & 0 & -\mathfrak{i}\lambda\delta_{k_y-}^{*} & -t\delta_{k_y+}^{*} & 0 & 0 \nonumber \label{Eqn: Lieb Ham K-Space}
\end{pmatrix} 
\end{eqnarray}

\noindent 
where $\delta_{k_i\pm}$=$1 \pm e^{\mathfrak{i}k_i}$, while $k_x$ and $k_y$ are the components of the momentum along the lattice vectors $\hat{a}_{x}$ and $\hat{a}_{y}$, respectively.  The basis used here is as in the dice lattice, i.e. $(c^{\phantom{dagger}}_{\mathbf{k},1,\uparrow}, \hspace{0.15cm} c^{\phantom{dagger}}_{\mathbf{k},1,\downarrow}, \hspace{0.15cm} c^{\phantom{dagger}}_{\mathbf{k},2,\uparrow}, \hspace{0.15cm} c^{\phantom{dagger}}_{\mathbf{k},2,\downarrow}, \hspace{0.15cm} c^{\phantom{dagger}}_{\mathbf{k},3,\uparrow}, \hspace{0.15cm} c^{\phantom{dagger}}_{\mathbf{k},3,\downarrow})$. $N_x$ and $N_y$ are the number of unit cells along the lattice vectors $\hat{a}_{x}$ and $\hat{a}_{y}$, respectively.

\subsection{Interacting Electrons in the Hartree-Fock Approximation}

To study the interaction effects, we added the onsite Hubbard repulsion term ($H_U=U\sum_{\mathbf{r},\alpha}n_{\mathbf{r},\alpha,\uparrow}n_{\mathbf{r},\alpha,\downarrow}$). This model cannot be solved exactly and in this study of interacting dice and Lieb lattices we used the standard Hartree-Fock (HF) decomposition in real space described as follows:
\begin{eqnarray}
&& \hspace*{3cm} H_{U} \approx  \nonumber \\
&& U\sum_{\substack{\mathbf{r},\alpha}}\left[\langle n_{\mathbf{r},\alpha,\uparrow}\rangle n_{\mathbf{r},\alpha,\downarrow} + \langle n_{\mathbf{r},\alpha,\downarrow}\rangle n_{\mathbf{r},\alpha,\uparrow} - \langle n_{\mathbf{r},\alpha,\uparrow}\rangle \langle n_{\mathbf{r},\alpha,\downarrow}\rangle \right. \nonumber\\
&& \left. -\lbrace\langle S^{+}_{\mathbf{r},\alpha} \rangle  S^{-}_{\mathbf{r},\alpha} + \langle S^{-}_{\mathbf{r},\alpha} \rangle  S^{+}_{\mathbf{r},\alpha} - \langle S^{+}_{\mathbf{r},\alpha} \rangle  \langle S^{-}_{\mathbf{r},\alpha}\rangle\rbrace \right] \label{Eqn: HF real-space}
\end{eqnarray}

\noindent
where $\langle n_{\mathbf{r},\alpha,\sigma} \rangle$ and $\langle S^{\pm}_{\mathbf{r},\alpha} \rangle$ are the charge and magnetic order parameters, respectively, for site $\alpha$ within the unit-cell $\mathbf{r}$ and spin projection $\sigma$. 

We simplified our HF results using that each unit cell in real space is a copy of all the rest, under the development of translationally invariant ferrimagnetic order, as found in Ref.~\cite{soni20}. Thus, $\langle n_{\mathbf{r},\alpha,\sigma} \rangle = \langle n_{\alpha,\sigma} \rangle$ and $\langle S^{\pm}_{\mathbf{r},\alpha} \rangle = \langle S^{\pm}_{\alpha} \rangle$. Under this condition, the interaction term in Eq.~\ref{Eqn: HF real-space} in momentum space becomes:
\begin{eqnarray}
\hspace*{-0.5cm}&&\hspace*{2cm} H_{U,\text{Quantum}} \approx \nonumber\\
\hspace*{-0.5cm}&& U\begin{pmatrix}
\langle n_{1,\downarrow}\rangle  & -\langle S^{-}_{1} \rangle & 0 & 0 & 0 & 0\\
-\langle S^{+}_{1} \rangle & \langle n_{1,\uparrow}\rangle & 0 & 0 & 0 & 0\\
0 & 0 & \langle n_{2,\downarrow}\rangle & -\langle S^{-}_{2} \rangle & 0 & 0 \\
0 & 0 & -\langle S^{+}_{2} \rangle & \langle n_{2,\uparrow}\rangle & 0 & 0  \\
0 & 0 & 0 & 0 & \langle n_{3,\downarrow}\rangle & -\langle S^{-}_{3} \rangle \\
0 & 0 & 0 & 0 & -\langle S^{+}_{3} \rangle & \langle n_{3,\uparrow}\rangle 
\end{pmatrix} \label{Eqn: HF Quantum in k-space} \\
\hspace*{-0.5cm}&&(H_{U,\text{Classical}})_{\alpha,\alpha} \approx U\left[\langle S^{+}_{\alpha} \rangle  \langle S^{-}_{\alpha}\rangle - \langle n_{\alpha,\uparrow}\rangle \langle n_{\alpha,\downarrow}\rangle \right] \label{Eqn: HF Classical in k-space}
\end{eqnarray}

\noindent
where $H_{U,\text{Quantum}}$ describes the quantum portion of the HF Hamiltonian and $H_{U,\text{Classical}}$ its classical component. Note that the basis for the interaction matrix in Eqs.~\ref{Eqn: HF Quantum in k-space} and  \ref{Eqn: HF Classical in k-space} is the same basis used for the non-interacting Hamiltonians in Eqs.~\ref{Eqn: Dice Ham K-Space} and \ref{Eqn: Lieb Ham K-Space}, i.e. $(c_{\mathbf{k},1,\uparrow}, \hspace{0.15cm} c_{\mathbf{k},1,\downarrow}, \hspace{0.15cm} c_{\mathbf{k},2,\uparrow}, \hspace{0.15cm} c_{\mathbf{k},2,\downarrow}, \hspace{0.15cm} c_{\mathbf{k},3,\uparrow}, \hspace{0.15cm} c_{\mathbf{k},3,\downarrow})$. More complicated orders, such as a spiral, would require the diagonalization of much larger matrices, but here a 6$\times$6 is sufficient to generate eigenvalues and eigenvectors.

Also, the presence of inversion symmetry with respect to the coordination-6 sites of the dice lattice and $C_{4}$ symmetry in the Lieb lattice helps us in reducing the number of order parameters. Under these symmetries the order parameters of the two blue sites within the unit cell must be the same.


To find the values of these order parameters, we performed self-consistent iterations derived from minimizing the Hamiltonian energy with respect to the mean-field parameters, while tuning the chemical potential accordingly to remain at the desired electronic density. In practice, we started with several random initial configurations (or seeds) for each order parameter (at fixed $U/t$ and $\lambda/t$) and inspected the lowest energy achieved after the iterative process. Then, we compared the ground-state energies from each of these converged results, and considered those with the lowest energy (sometimes the results of
different iterative processes lead to different energies due to trapping in metastable states, thus the importance of using
a variety of initial random order parameter sets). 

More specifically, to reach the self-consistent solution in the Hartree-Fock order parameters, we used the simple mixing as described below:
\begin{equation}
|O^{n+1}_{in}\rangle = (1-\alpha)|O^{n}_{in}\rangle + \alpha|O^{n}_{out}\rangle,
\end{equation}
where $|O^{n}_{in}\rangle$ is the input array of order parameters for the $n$-th iteration and $|O^{n}_{out}\rangle$ is calculated using the eigenspectrum of the Hartree-Fock Hamiltonian for the given density of electrons~\cite{mixings}. The chemical potential is tuned to reach the targetted electronic density, in this case half-filling, 
for a fixed very low temperature of $T=0.0001t$. We used $\alpha=0.5$ in the previous equation.
The convergence error criterium of our HF results was $10^{-6}$. 

Finally, in the Appendix we show evidence that using the full Hartree-Fock approximation, as opposed to only Hartree, in the cases of the dice and Lieb lattices is qualitatively important. Not only the energies are better with HF, but in addition, the Chern numbers are different than those obtained when only using Hartree, indicating that the Fock terms are relevant.


\begin{figure}[!t] 
\centering
\includegraphics[width=3.4in, height=2.8in]{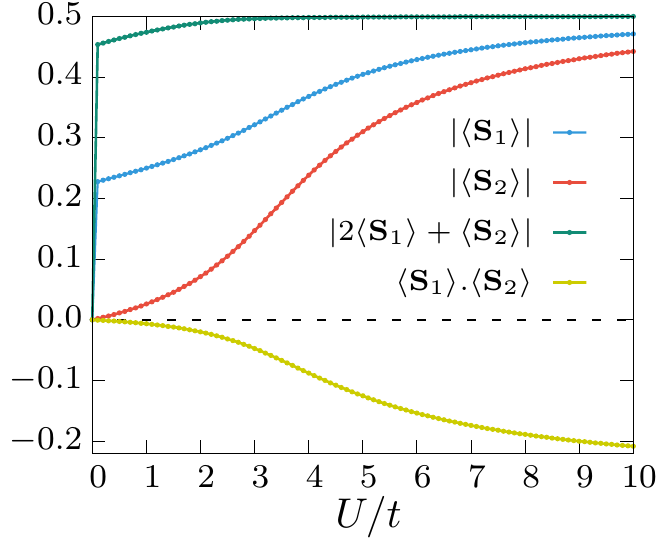}
\caption{Average spin moments vs $U/t$ at $\lambda=0.3t$ and $\epsilon=0.6t$ via Hartree-Fock at half-filling on a 60$\times$60 unit-cell system used for the self-consistency. $|\langle \mathbf{S}_1 \rangle|$ and $|\langle \mathbf{S}_2 \rangle|$ are the magnitude of the spins at sites 1 and 2, respectively and $|2\langle \mathbf{S}_1 \rangle + \langle \mathbf{S}_2 \rangle|$ is the net spin moment of the unit cell. $\langle \mathbf{S}_1 \rangle . \langle \mathbf{S}_2 \rangle$ shows the dot product of the spins at sites 1 and 2. The absolute values just denote the fact that depending on initial seeds the overall order parameter can be positive or negative with equal chance, as in any ferro or ferri system, but the smoothness of the results shows that convergence was properly achieved even using different seeds at each point. 
\label{Fig: FMO vs U at SOC=0.3 via HF}}
\end{figure}

\section{Results}
In this section, we will discuss the Hartree-Fock results for the two-dimensional (2D) dice and Lieb lattices. Surprisingly, we observed many different topological phases and present them in our phase diagrams for both respective lattices, see Figs.~\ref{Fig: Phase Diagram Dice} and \ref{Fig: Phase Diagram Lieb}. Each state in these phase diagrams is characterized by the set of Chern numbers $(C_1,C_2,C_3,C_4,C_5,C_6)$ calculated for each of the six bands, from the bottom up increasing energy, arising from the 6$\times$6 diagonalization of the matrices shown in the previous section after convergence, all at half-filling. We also observed that all topological phase transitions in our systems occur through a band touching point, as expected for topological phase transitions. Namely, varying a parameter such as $U$ or $\lambda$, first a gap exists among all phases, then at one point a gapped region between two bands becomes gapless when those two bands touch,  and then the gap reopens again, with a concomitant change in the Chern numbers of the two bands involved. Concrete examples are shown below.

\subsection{Dice Lattice Results}
We start by considering a 60$\times$60 unit cell dice lattice system, with 60 unit cells along each lattice vectors $\hat{e}_1$ and $\hat{e}_2$, see Fig.~\ref{Fig: Geometries}(a) for reference. We study the ground state properties of the dice Hamiltonian on this lattice in the presence of interactions at half-filling, via Hartree-Fock. When $U/t=0$, degenerate flat bands are present at $E=0$ in this lattice even for $\lambda/t$ nonzero~\cite{soni20,wang11} (the same occurs for the Lieb
lattice shown below). For any finite $U$, these flat bands split into two non-degenerate bands around $E=0$, even in the absence of external fields. Also, with the inclusion of $U/t$ long-range ferrimagnetic order develops in the system. Our previous DMRG+Lanczos study showed the presence of this ferrimagnetic order for $N\times 2$ ribbons of dice lattice~\cite{soni20}. However, to confirm that this type of order dominates also in the present 2D case, i.e. not just in ribbons, a comprehensive study of the magnetic properties was carried out via HF.


In Fig.~\ref{Fig: FMO vs U at SOC=0.3 via HF}, we show that the ordering of the local spins are indeed ferrimagnetic using HF. Firstly, the magnitude of the spin at site 2 ($|\langle \mathbf{S}_2 \rangle|$), i.e. the red sites, is always smaller as compared to the magnitude of the spin at site 1 ($|\langle \mathbf{S}_1 \rangle|$), i.e. the blue sites. In addition, the product of the two spins $\langle \mathbf{S}_1 \rangle$ and $\langle \mathbf{S}_2 \rangle$ is always negative. Moreover, at any finite $U/t$ both $\langle \mathbf{S}_1 \rangle$ and $\langle \mathbf{S}_2 \rangle$ are collinear: we verified that $\langle \mathbf{S}_1 \rangle.\langle \mathbf{S}_2 \rangle/|\langle \mathbf{S}_1 \rangle||\langle \mathbf{S}_2 \rangle|=-1$. With all this information, we can safely conclude that the ordering of the spins in our 2D dice lattice is ferrimagnetic, as conjectured in Ref.~\cite{soni20} studying small clusters.

\begin{figure}[!b] 
\centering
\includegraphics[width=3.4in, height=2.8in]{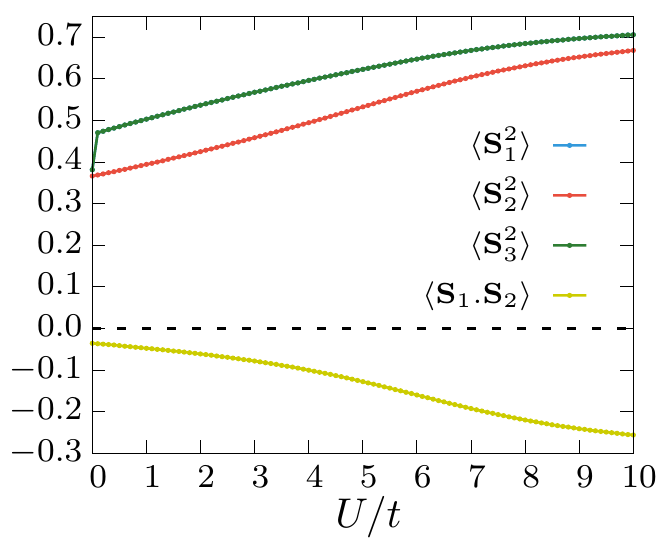}
\caption{Average local moments $\langle \mathbf{S}_\alpha^2\rangle$ and spin-spin correlation $\langle \mathbf{S}_1 . \mathbf{S}_2 \rangle$ vs $U/t$ at $\lambda=0.3t$ and $\epsilon=0.6t$, obtained via Lanczos at half-filling on a 2$\times$2 unit-cell system. The results for site 1 merely confirm that by symmetry sites 1 and 3 must behave identically. This is why only green is observed in the figure, while blue is hidden behind. \label{Fig: FMO vs U at SOC=0.3 via Lanczos}}
\end{figure}

At finite $U/t$, we have not observed any further magnetic transition in Fig.~\ref{Fig: FMO vs U at SOC=0.3 via HF} and the magnetic ordering is consistently ferrimagnetic for the entire range of $U/t$. However, there is an abrupt change in the magnetic ordering when moving from $U/t=0$ to $U/t=0.1$, the first point studied after $U/t=0$ in our grid of points, where there is a sudden jump in the magnitude of $\langle \mathbf{S}_1 \rangle$. This is because the flat band at $U/t=0$ consists of states from coordination-3 sites and even a small value of $U/t$ breaks the global degeneracy that causes the $E=0$ flat band, leading to a jump in $|\langle \mathbf{S}_1 \rangle|$. In other words, the sudden split of the flat band separates that original band into two, each with a different orientation of the ferri order parameter. To confirm these results, we performed Lanczos on a 2$\times$2 system, see Fig.~\ref{Fig: FMO vs U at SOC=0.3 via Lanczos}, where we observed the same features being captured in the average local moments $\langle \mathbf{S}_1^2\rangle$. Here we also show that, as expected by mere symmetry even with the quantum fluctuations incorporated, $\langle \mathbf{S}_1^2\rangle$ and $\langle \mathbf{S}_3^2\rangle$ are identical to one another.

\begin{figure}[!t] 
\hspace*{-0.5cm}
\includegraphics[width=3.5in, height=3.5in]{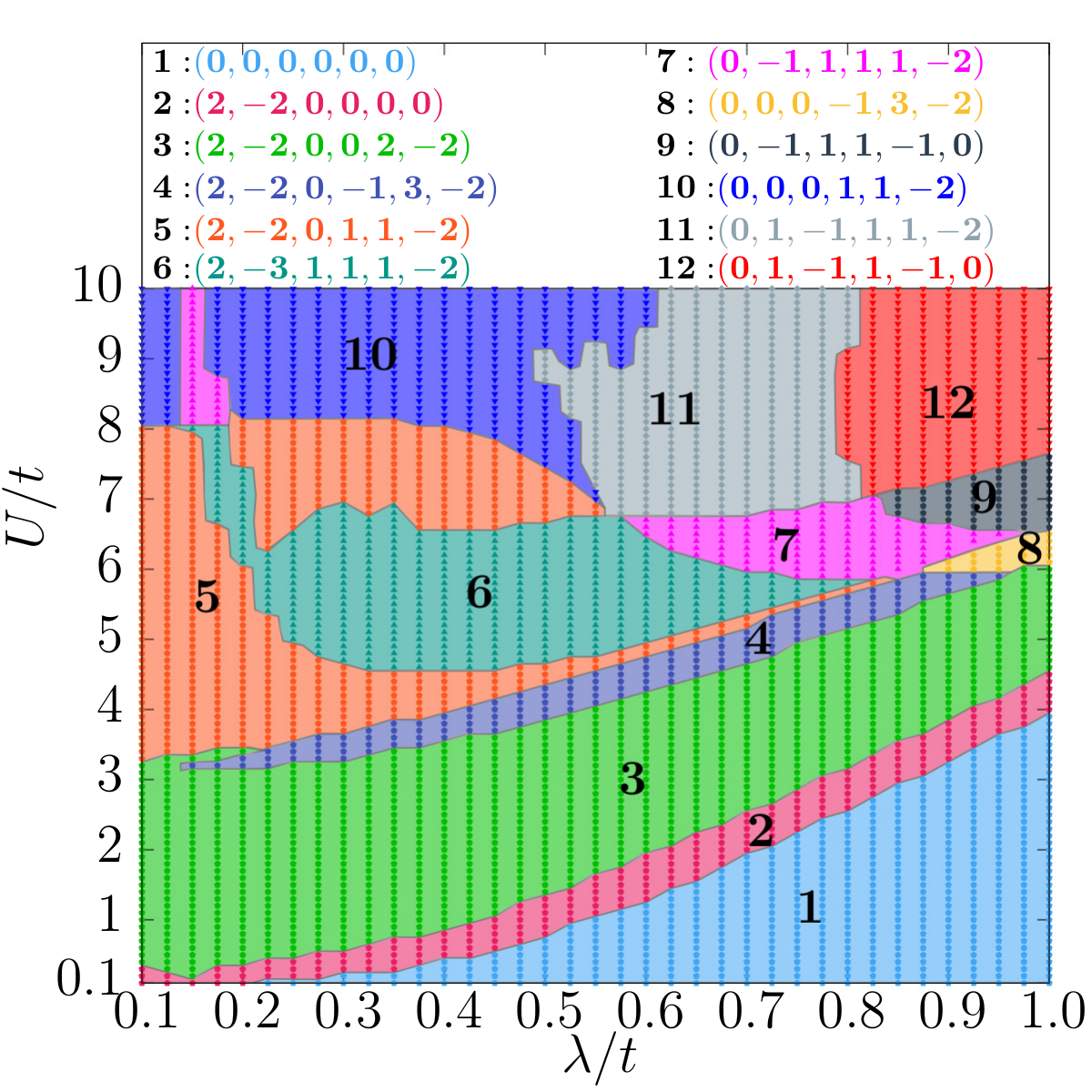}
\caption{$U/t$ vs $\lambda/t$ phase diagram for the 60$\times$60 dice lattice,
calculated using the Hartree-Fock approximation. $\epsilon=0.6t$ is used here. The different colors refer to different sets of Chern numbers $(C_1,C_2,C_3,C_4,C_5,C_6)$, according to the color convention indicated at the top. Note that all phases
are ferrimagnetic including the $(0,0,0,0,0,0)$ phase 1. 12 different topological phases were identified. 
\label{Fig: Phase Diagram Dice}}
\end{figure}

In Fig.~\ref{Fig: Phase Diagram Dice}, we display the $U/t$ versus $\lambda/t$ topological phase diagram for a 60$\times$60 unit cells dice lattice system, at $\epsilon=0.6t$. To establish this phase diagram, we computed the first order Chern number of all the 6 non-degenerate bands that arise from the HF approximation, at half-filling, using the method introduced in Ref.~\cite{fukui05}, involving individual plaquettes in the discretized grid in momentum space of the lattice used. Unlike in the case of the Lieb lattice, where the poles of the Berry curvature lie at the boundary of the first Brillouin zone, as described later in the text, in the dice lattice they lie well within the first Brillouin zone. Hence, the calculation of Chern numbers here is quite straightforward.

It is intersting to note that for $\epsilon=0$ the Hamiltonian in equation \ref{Eqn: NI Ham Real Space} is invariant, for half-filling, under the particle-hole transformation shown below:
\begin{equation}
c_{\mathbf{r},\alpha,\sigma} \rightarrow \nu_{\sigma} e^{i\pi\alpha}  c^{\dagger}_{\mathbf{r},\alpha,\bar{\sigma}}
\end{equation}
where $\nu_{\sigma}=1(-1)$ for $\sigma=\uparrow(\downarrow)$. The presence of $\epsilon \neq 0$ breaks the particle-hole symmetry which leads to asymmetry in the Chern numbers i.e. $C_{i}\neq -C_{6-i}$, as noticed in many phases in the phase diagram.

\begin{figure}[!t] 
\centering
\includegraphics[width=3.4in, height=4in]{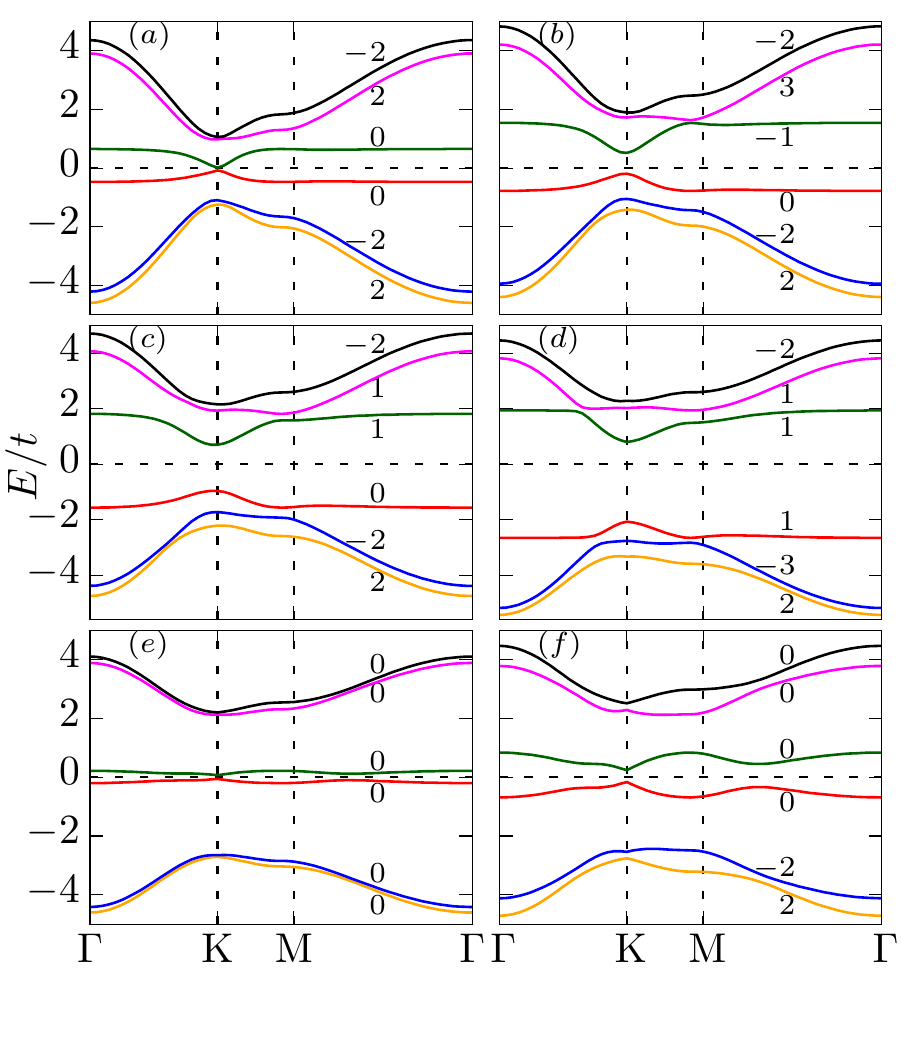}
\vspace*{-1cm}
\caption{Panels $(a), (b), (c)$ and $(d)$ represent the bands at $\lambda=0.3t$ for the phases $(2,-2,0,0,2,-2)$, $(2,-2,0,-1,3,-2)$, $(2,-2,0,1,1,-2)$ and $(2,-3,1,1,1,-2)$, at  $U/t=2.0$, $U/t=3.4$, $U/t=4.4$  and $U/t=5.5$, respectively. Panels $(e)$ and $(f)$ represent the bands at $\lambda/t=0.8$ corresponding to the phases $(0,0,0,0,0,0)$, and $(2,-2,0,0,0,0)$ at $U/t=1.0$, and $U/t=3.0$, respectively. All the plots were obtained using a 60$\times$60 grid in momentum space, and $\epsilon=0.6t$. The numbers next to each band are the Chern numbers of those bands. \label{Fig: Multiple Phases Dice}}
\end{figure}

As illustration, in Fig.~\ref{Fig: Multiple Phases Dice}, we are showing representative bands for some of the phases that appear in our phase diagram. We observed that for small values of the interaction strength $U/t$, there are three different classes of bands that are isolated in pairs (as expected from continuity starting at $U/t=0$ where there are three bands, each with degeneracy two). For example, in Figs.~\ref{Fig: Multiple Phases Dice}$(a)$ and \ref{Fig: Multiple Phases Dice}$(f)$ involving the phases $(2,-2,0,0,2,-2)$ and $(2,-2,0,0,0,0)$, the lower two bands, the middle two bands, and the upper two bands form classes of their own and each has a net sum of Chern numbers equal to zero. Increasing $U/t$, the middle two bands split further and now we have two different classes made of three lower and three upper bands, see Fig.~\ref{Fig: Multiple Phases Dice}$(b)$, \ref{Fig: Multiple Phases Dice}$(c)$, and \ref{Fig: Multiple Phases Dice}$(d)$ that represent the phases $(2,-2,0,-1,3,-2)$, $(2,-2,0,1,1,-2)$ and $(2,-3,1,1,1,-2)$ respectively. Note now the net sum of Chern numbers of the lower three bands and upper three bands is zero separately. This last issue is worth remarking: in the dice lattice at half-filling, our results predict that the three lower bands have Chern numbers that {\it always} add up to zero in the entire phase diagram, suggesting that the Anomalous Quantum Hall Effect (AQHE) will cancel. 
However, in the Lieb lattice, as shown below, this situation will only occur in a fraction of the phase diagram.

\begin{figure}[!b] 
\centering
\includegraphics[width=3.4in, height=2.8in]{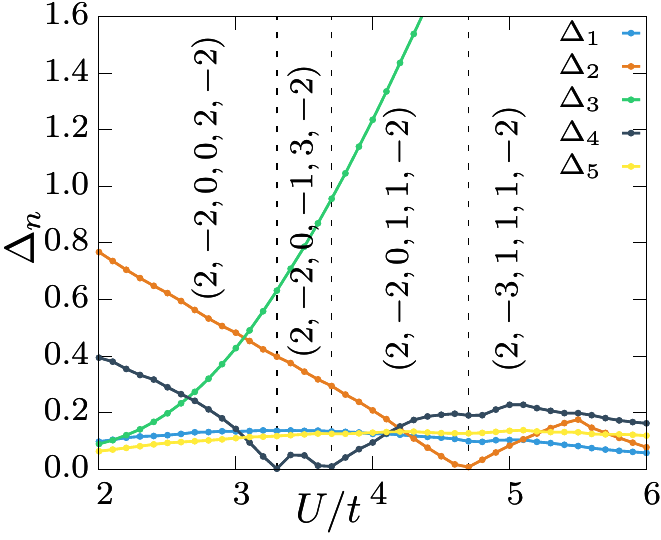}
\caption{Band gaps $\Delta_n$ vs $U/t$ plot for $\lambda/t=0.3$ and $\epsilon/t=0.6$ via Hartree-Fock at half-filling on a 60$\times$60 unit-cell system. We show $\Delta_n=\min_{\mathbf{k}} \left[E_{n+1}(\mathbf{k})-E_{n}(\mathbf{k})\right]$, where $n$ is the band index. Here we can clearly observe the topological transition points around $U/t \sim 3.3$, $U/t \sim 3.7$, and $U/t \sim 4.7$, respectively. \label{Fig: Band Gaps vs U at SOC=0.3}}
\end{figure}

We observed that to characterize the topological phase transitions and find the precise locations of the transitions, the magnetic observables, such as the ferri order parameter, are certainly insufficient. For example, we did not detect any
noticeable modification in the first and second derivatives of the ferrimagnetic order parameters vs $U/t$. 
This is in agreement with the transitions being topological.
Thus,  we calculated $\Delta_n$ which is the minimum gap in energy between the $n^{th}$ and $n+1^{th}$ energy bands at fixed SOC $\lambda=0.3t$ as example, as in Fig.~\ref{Fig: Band Gaps vs U at SOC=0.3}. Here we show that whenever a topological phase transition occurs the bands go through a band touching point, namely $\Delta_n=0$. As example of this behavior, we will consider the phase transition occuring between the phases
$(2,-2,0,0,2,-2)$ and $(2,-2,0,-1,3,-2)$. In this case, the Chern numbers of the band 4 and band 5 change from $(0,2)$ to $(-1,3)$ which implies that somewhere between these two phases there should be a value of $U/t$, at fixed $\lambda/t$, where $\Delta_4=0$. In Fig.~\ref{Fig: Band Gaps vs U at SOC=0.3}, we can see that $U/t\sim 3.3$ corresponds to that touching point, confirming the topological nature of the transitions.

\begin{figure}[!t] 
\centering
\includegraphics[width=3.4in, height=6in]{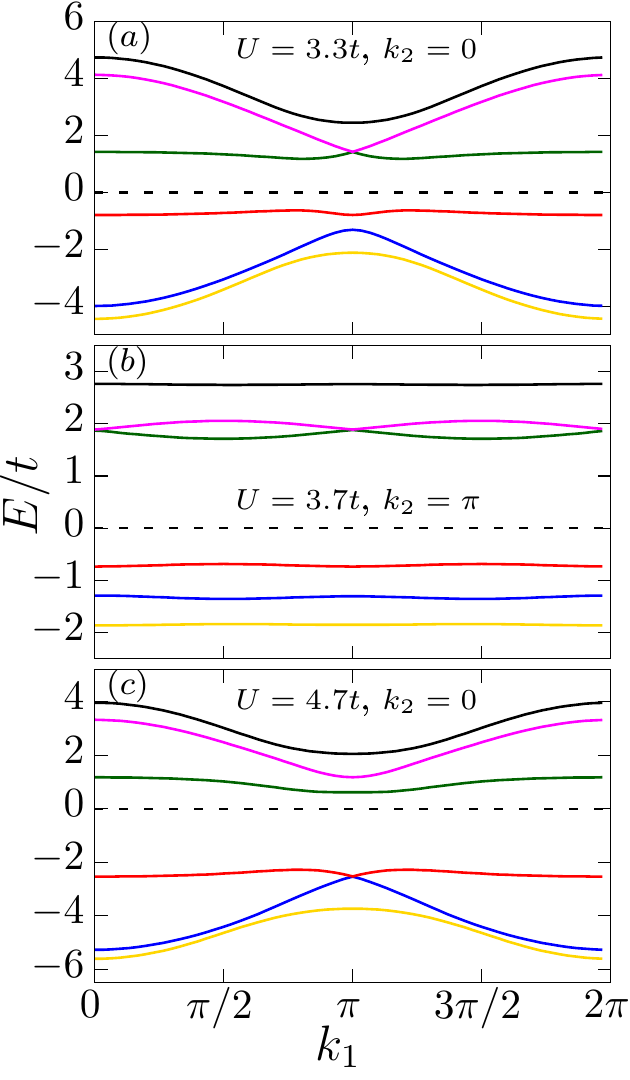}
\caption{Energy bands vs $k_1$ plots (for the specific values of $k_2$'s that highlight the band touching region) at the transition coupling values $U/t = 3.3, 3.7$ and $4.7$, and at $\lambda/t=0.3$ and $\epsilon/t=0.6$. via Hartree-Fock on a 60$\times$60 unit-cell system. Panel (a) represents the transition point from phase $(2,-2,0,0,2,-2)$ to phase $(2,-2,0,1,3,-2)$, whereas plot (b) represents the transition point from phase $(2,-2,0,1,3,-2)$ to phase $(2,-2,0,1,1,-2)$. Lastly, plot (c) represents the transition point from  phase $(2,-2,0,1,1,-2)$ to phase $(2,-3,1,1,1,-2)$. \label{Fig: Phase Transition at SOC=0.3}}
\end{figure}

Similarly, we show in detail two more such transition points for $\lambda/t=0.3$ at $U/t \sim 3.7$, and $U/t \sim 4.7$,  where $\Delta_4=0$ when $(2,-2,0,-1,3,-2) \rightarrow (2,-2,0,1,1,-2)$ and $\Delta_2=0$ when $(2,-2,0,1,1,-2) \rightarrow (2,-3,1,1,1,-2)$, respectively. Also, we noticed that for a specific band $n$ while moving from one phase to another a net change in Chern number of $|\Delta C_{n}|=1$ or $2$ is observed in the dice lattice. For example, in Fig. \ref{Fig: Band Gaps vs U at SOC=0.3} while moving from phase $(2,-2,0,0,2,-2)$ to $(2,-2,0,-1,3,-2)$ we observe $|\Delta C_{4}|=|\Delta C_{5}|=1$. Similarly, from phase $(2,-2,0,-1,3,-2)$ to $(2,-2,0,1,1,-2)$ we observe $|\Delta C_{4}|=|\Delta C_{5}|=2$. This is true for all the phase transitions in our phase diagram in Fig.~\ref{Fig: Phase Diagram Dice}.

In Fig.~\ref{Fig: Phase Transition at SOC=0.3}, we have plotted the bands associated with the three transition points reported in Fig.~\ref{Fig: Band Gaps vs U at SOC=0.3}. At finite $U/t$, the symmetry points $\Gamma$, $\mathrm{K}$ and $\mathrm{M}$ are not necessarily the location of the bands touchings, although in practice they turned out to be. Hence, a complete Brillouin zone check is in principle required. For that purpose, we plotted the bands versus the lattice momentum $k_1$, for different values of $k_2$'s. In Fig.~\ref{Fig: Phase Transition at SOC=0.3}$(a)$, we depict the band touching point at $U/t=3.3$. This band touching point here lies at momentum $(k_1,k_2)=(\pi,0)$ and is present between bands 4 and band 5. Similarly, in Fig.~\ref{Fig: Phase Transition at SOC=0.3}$(b)$ and \ref{Fig: Phase Transition at SOC=0.3}$(c)$, we explicitly show the band touching points for the transition values $U/t=3.7$ and $U/t=4.7$, respectively. For $U/t=3.7$, the band touching point lies at the momentum point $(\pi,\pi)$ and occurs between bands 4 and 5, whereas for $U/t=4.7$ the touching lies at momentum $(\pi,0)$ and is present between bands 2 and 3.

\begin{figure}[!b] 
\centering
\includegraphics[width=3.4in, height=2.8in]{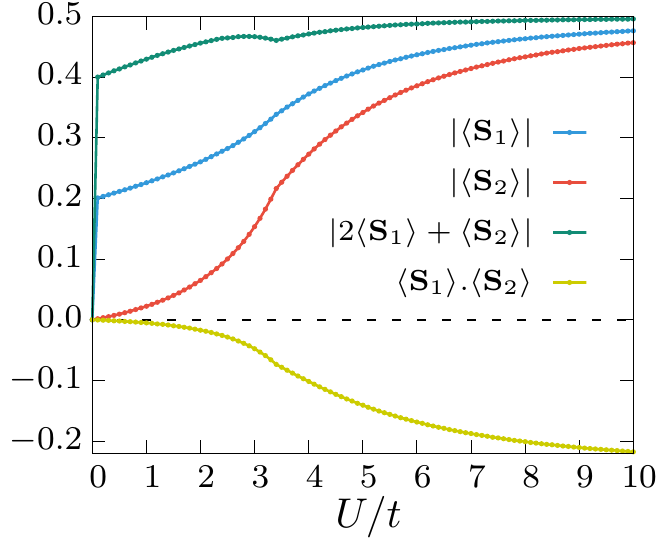}
\caption{Average spins vs $U/t$, at $\lambda/t=0.45$ and $\epsilon/t=0.5$ obtained via Hartree-Fock at half-filling on a 64$\times$64 unit-cell Lieb lattice system. The $U/t=0$ jumps occur for the same reason as in the dice lattice, namely the splitting of the $E=0$ flat band immediately when turning on $U/t$.  \label{Fig: FMO vs U for Lieb at SOC=0.45 via HF}}
\end{figure}

\subsection{Lieb Lattice Results}

Let us now discuss our HF results for the Lieb lattice~\cite{lieb-original}. Similar to the  dice lattice, here we start by considering a two-dimensional 64$\times$64 system, with 64 unit cells along each lattice vectors $\hat{a}_x$ and $\hat{a}_y$ (readers are referred to Fig.~\ref{Fig: Geometries}$(b)$ for the geometry). The non-interacting properties of the Lieb lattice entail degenerate flat bands at half-filling at $E=0$~\cite{chen17}, as in the case of the dice lattice. Also as in the dice lattice, we have observed that after the inclusion of the onsite Hubbard $U/t$ the flat band immediately splits into two non-degenerate bands. However, unlike the dice lattice, the splitting of the Lieb flat band adds unexpected technical complications because special points in momentum space remain very close to one another, even after the splitting induced by $U/t$ and $\lambda/t$. Thus, considerably more 
numerical effort is required to make sure true gaps are formed in the Lieb lattice than in the dice lattice.


\begin{figure}[!t] 
\hspace*{-0.5cm}
\includegraphics[width=3.5in, height=3.5in]{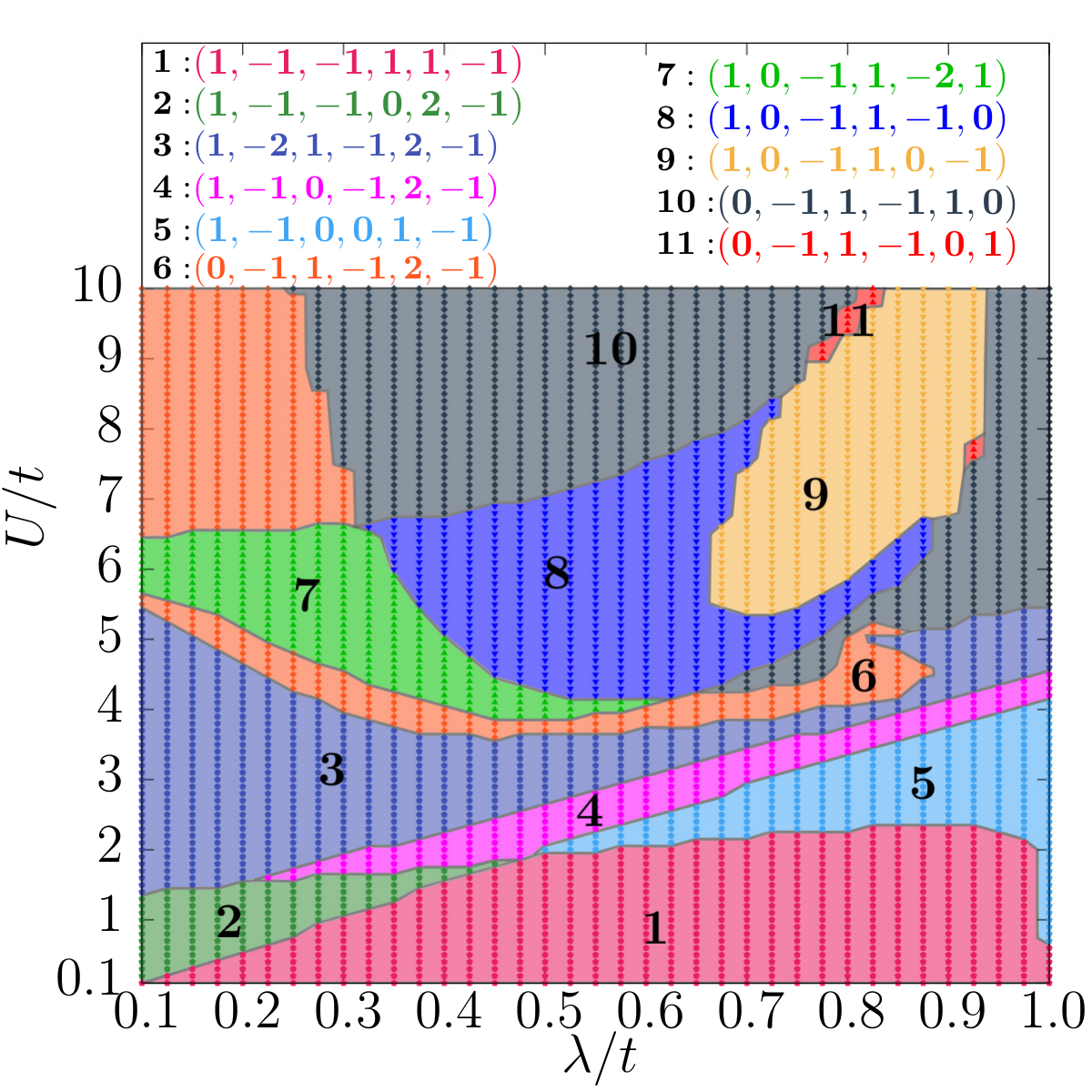}
\caption{$U/t$ vs $\lambda/t$ phase diagram for a 64$\times$64 Lieb lattice,
calculated using the Hartree-Fock approximation. $\epsilon/t=0.5$ was used here. The color convention and its relation with the Chern numbers of the bands, from bottom to top in energy, is shown at the top of the figure.  \label{Fig: Phase Diagram Lieb}}
\end{figure}

In Fig.~\ref{Fig: FMO vs U for Lieb at SOC=0.45 via HF}, we illustrate the magnetic properties of the ground state via HF at half-filling. We followed the same procedure mentioned before in the dice lattice section, and again we concluded that the 2D Lieb lattice at finite $U/t$ also exhibits ferrimagnetism. We found that the magnitude of the spin at sites 2 ($|\langle \mathbf{S}_2 \rangle|$) is always smaller as compared to the magnitude of the spin at sites 1 ($|\langle \mathbf{S}_1 \rangle|$), while the dot product of the two spins $\langle \mathbf{S}_1 \rangle .\langle \mathbf{S}_2 \rangle$ is always negative. Moreover, all the spins are always collinear. Then, this information helps us to establish that the ground-state for the Lieb lattice is ferrimagnetic as for the dice lattice. However, unlike the dice lattice, here we have observed a magnetic anomaly at $U\sim 3.4t$. For example, see the change in slope in the $|2\langle \mathbf{S}_1 \rangle + \langle \mathbf{S}_2 \rangle|$ curve in Fig.~\ref{Fig: FMO vs U for Lieb at SOC=0.45 via HF}. However, it does not influence on the symmetry breaking pattern, nor on the prediction of topological phase transitions in our system. The origin of this strange anomaly will be studied in future work, and its presence is not crucial for the discussion that follows.


For the case of the Lieb lattice the method of Ref.~\cite{fukui05} to calculate Chern numbers
may have problems because the singular portions of the 
Berry curvature that contribute to the Chern number are located at the boundary of the first BZ. We have found two solutions to this problem:

(1) The Lieb lattice contains 3 sites $(1,2,3)$ (see Fig.~\ref{Fig: Geometries}(b)) 
per unit cell. Each site has one active orbital and, as a result, we have an effective three-orbital model, 
although the three orbitals have different locations in the unit cell. Thus, the wave function is not periodic 
in the first Brillouin zone (BZ). The periodicity instead is of two BZs in each direction ($x$ and $y$). Thus, we can calculate the Chern number by focusing on an extended, instead of single unit, BZ with momentum $k_i$ in the interval $[0,4\pi)$ instead of $[0,2\pi)$ for both directions $i=x,y$. In this situation, the admissibility condition for the calculation, 
described in Ref.~\onlinecite{fukui05} is now satisfied.

(2) However, there is another procedure that leads to the same results: using a gauge transformation 
will allow us to evaluate the Chern number in a single BZ. This gauge transformation 
effectively places the three orbitals at the same site, i.e. it maps sites 1 and 3 into site 2, 
restoring the periodicity of the wave function.

The gauge transformation is given by:
\begin{equation}
U=\left(\begin{array}{ccc}
e^{ik_x/2} & 0 & 0  \\
0 & 1  & 0  \\
0 & 0 & e^{ik_y/2}
\end{array} \right)
\label{eqsuppl}
\end{equation}

\noindent for the sites $(1,2,3)$ as in Fig.~\ref{Fig: Geometries}(b). Defining $H'=UHU^{-1}$ we can calculate the Chern number 
in the traditional way since the wavefunctions are now periodic in the  first BZ and the admissibility condition described in Ref.~\cite{fukui05} is now satisfied. This transformation is similar in spirit to the approach in Ref.~\cite{ruiyu} to evaluate 
the $Z_2$ topological invariant for band insulators.
We have verified that the Chern numbers are identical using both methods (1) and (2). 
The second approach reduces the number of points in $k$-space needed to compute the Chern numbers.


In Fig.~\ref{Fig: Phase Diagram Lieb}, the $U/t$ vs $\lambda/t$ topological phase diagram for the Lieb lattice in the HF approximation is displayed, at $\epsilon=0.5t$. As for the dice lattice, here we computed the first order Chern number of all the 6 non-degenerate bands at half-filling. For the Lieb lattice, we used the methods (1) and (2) described above in momentum space to verify consistency in many points,
but primarily the methodology (1). As for the case of the dice lattice, the plethora of topological phases is remarkable, with 11 of them, all displaying ferrimagnetic order. Previous studies of non-interacting electrons with Rasba coupling, 
using in addition staggered magnetic fields which qualitatively resemble the ferrimagnetic order, also reported a rich
phase diagram but with only 4 different topological phases~\cite{chen17}. Other studies of the Lieb lattice using non-interacting
electrons with variations of the real next-nearest-neighbor hopping~\cite{morais} also reported rich topological phase diagrams. 

As mentioned for the dice lattice where $|\Delta C_{n}|=1$ or $2$ is observed as a phase transition condition, in the Lieb lattice we noticed that for a specific band $n$ while moving from one phase to another a net change in Chern number of $|\Delta C_{n}|=1$ is observed. This is true for all the phase transitions in our Lieb lattice phase diagram in Fig.~\ref{Fig: Phase Diagram Lieb}.



\begin{figure}[!t] 
\centering
\includegraphics[width=3.4in, height=3.2in]{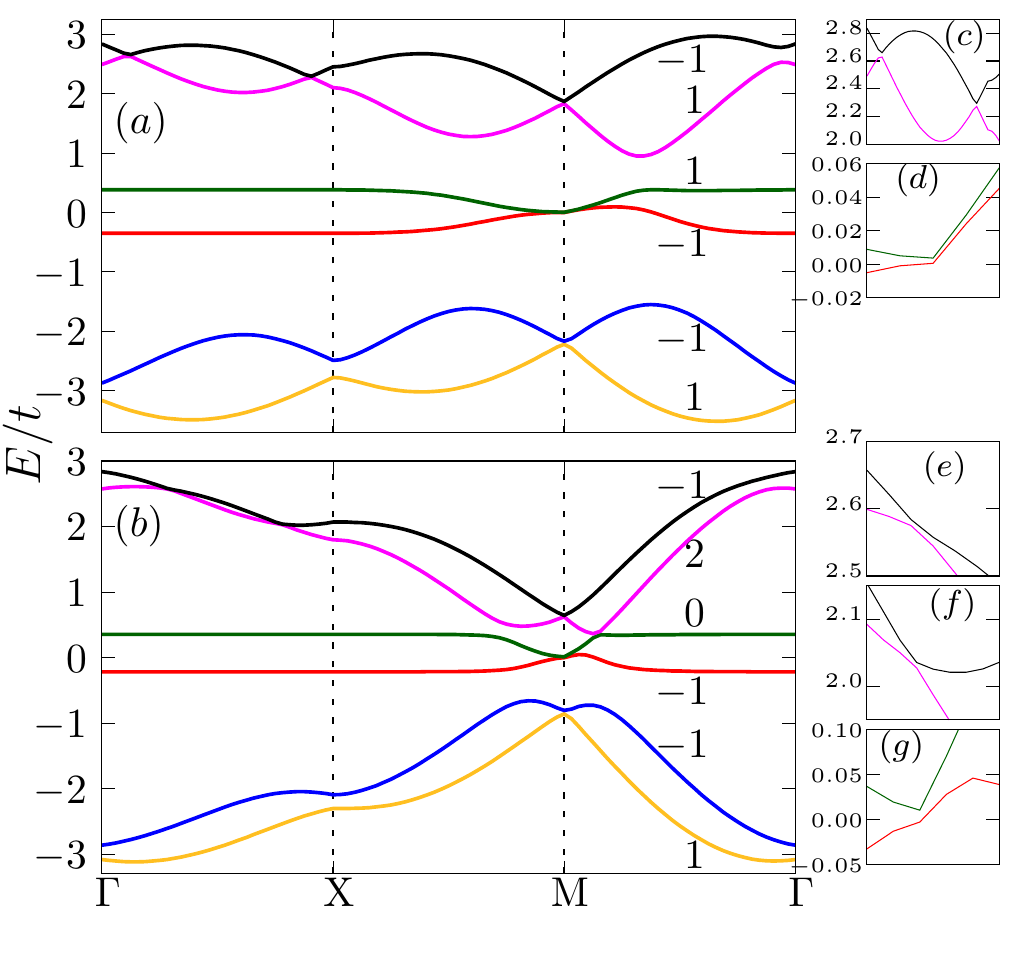}
\caption{Representative bands for some phases shown in the phase diagram at half-filling. Panel $(a)$ is for $U/t=1.7$ and $\lambda/t=0.7$, representing the phase $(1,-1,-1,1,1,-1)$. Panel $(b)$ is for $U/t=1.1$ and $\lambda/t=0.225$, representing the phase $(1,-1,-1,0,2,-1)$. The insets panels (c) and (d) amplify points where bands are very close to one another, illustrating that there is an abnormally small but nonzero finite gap between the top two and middle two bands, respectively. The gap between the two bands at the bottom is already visible in panel (a). Inset plots $(e)$ and $(f)$ depicts the finite gap between the top two bands, whereas $(g)$ shows the finite gap between the middle two bands. Again, the gaps for the two bottom lines are already visible in panel (b). All the plots are for a 64$\times$64 Lieb lattice, calculated using the Hartree-Fock approximation. $\epsilon/t=0.5$ is used here. \label{Fig: CN Non-Zero Band Plots}}
\end{figure}

\begin{figure}[!t] 
\centering
\includegraphics[width=3.2in, height=5.5in]{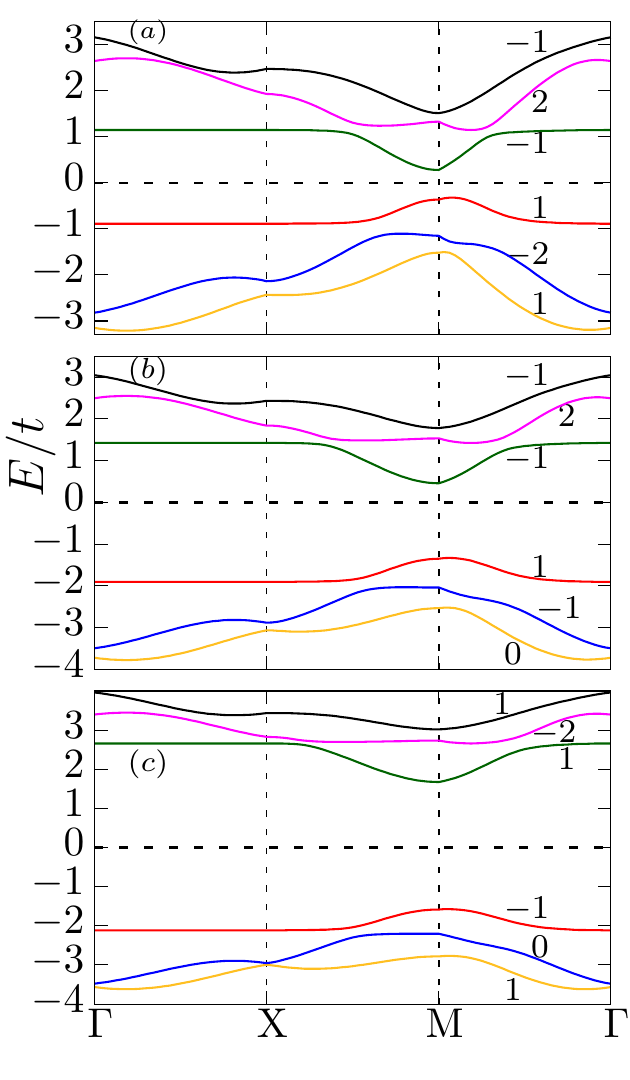}
\caption{Representative bands for some typical phases at half-filling in the phase diagram. Panels $(a)$, $(b)$, and $(c)$ are for $U/t=3.0$, $U/t=4.2$, and $U/t=5.5$, respectively, using parameters $\lambda/t=0.3$ and $\epsilon/t=0.5$. $(a)$ contain the bands from the phase $(1,-2,1,-1,2,-1)$, $(b)$ represents the bands from the phase $(0,-1,1,-1,2,-1)$, while $(c)$ are the bands from the phase $(1,0,-1,1,-2,1)$. All the plots are for a 64$\times$64 Lieb lattice, calculated using the Hartree-Fock approximation. \label{Fig: CN Zero Band Plots}}
\end{figure}

In Figs.~\ref{Fig: CN Non-Zero Band Plots}$(a)$ and \ref{Fig: CN Non-Zero Band Plots}$(b)$, we display representative bands for the phases $(1,-1,-1,1,1,-1)$ and $(1,-1,-1,0,2,-1)$, as example, in the weak coupling regime where the original group of three bands of $U/t=0$ (each doubly degenerate) can still be observed.
At first impression in the scale used, the bands seem to have band touching points, namely the abnormally small gaps in this band structure are not visible to the eye. To show that actually there is a tiny but nonzero gap in our results we have included some insets where by changing the scale, using a finer grid of points, and focussing on the apparent touching points, we show that small gaps are actually present between these curves (see insets plots \ref{Fig: CN Non-Zero Band Plots}$(c)$ to \ref{Fig: CN Non-Zero Band Plots}$(g)$). Similarly small gaps were reported before in Ref.~\cite{chen17} for the same Lieb lattice but in a staggered magnetic field. An important qualitative observation is that if we add up the  Chern number of the lowest three bands, namely those populated at half-filling, they add to a nonzero Chern number and as a consequence an AQHE is to be expected, similarly as it
occurs for the dice lattice but in a uniform magnetic field at $U/t=0$~\cite{wang11}, instead of the ferrimagnetic order found here.

In Fig.~\ref{Fig: CN Zero Band Plots}$(a)$, \ref{Fig: CN Zero Band Plots}$(b)$ and \ref{Fig: CN Zero Band Plots}$(c)$, we continue showing representative bands for the phases $(1,-2,1,-1,2,-1)$, $(0,-1,1,-1,2,-1)$ and $(1,0,-1,1,-2,1)$, respectively, at larger values of $U/t$. Because of the large $U/t$, and as in the dice lattice, three bands are now grouped together at low energies and three at high energies.
Note that in this figure, if we fill with electrons up to half-filling, the sum of Chern numbers is now {\it zero}, and as a consequence no AQHE is expected. Thus, although not a phase transition, there are two regimes in the Lieb phase diagram, one with AQHE nonzero and one with AQHE zero at half-filling, adding an extra interesting detail to our results. It is remarkable that nonzero AQHE does not occur in any of the phases of the dice lattice: this is the only, but important, difference we found between the dice and Lieb lattice that otherwise behave very similarly within the HF approximation, both with many topological phases.


\section{Conclusions}

The simultaneous study of the effect of Hubbard correlation and spin-orbit coupling in electronic models is widely considered among the most important next challenges in condensed matter theory.
In this publication, we presented the phase diagrams of the dice and the Lieb lattices, including Rashba spin-orbit coupling and Hubbard onsite repulsion, within the Hartree-fock approximation to treat electronic correlation effects.

A surprisingly rich phase diagram was unveiled in both cases. While regarding canonical spontaneous symmetry breaking both lattices display the same ferrimagnetic order, as predicted for the dice case in Ref.~\cite{soni20} using small cluster Lanczos, our present work unveiled a plethora of ``hidden'' topological transitions where the Chern numbers of the bands change at the boundaries between phases. In these topological transitions, gaps between pairs of bands close and reopen varying parameters, and before and after the closing the resulting Chern numbers are different. The abundance of phases is surprising: without calculating the Chern numbers, {\it a priori} it would have been impossible to anticipate that topological transitions occur because finding the exact place where the closing of the gap occurs is in principle quite difficult (we showed a few examples). The regions of zero gap are a web-like manifold of dimension 1 in the dimension 2 of the phase diagram varying Hubbard $U/t$ and Rashba $\lambda/t$ couplings, at a fixed onsite energy $\epsilon/t$ difference between sites with different coordination number. 

Moreover, as already expressed, the entire phase diagram is ferrimagnetic, and this order parameter appears to behave smoothly across the topological phase transitions. This confirms via a toy model the growing perception in the community that topology is ``everywhere'', namely that a large percentage of materials studied for years in fact have nontrivial topological properties. The same seems to occur with seemingly ``harmless'' models of interacting electrons, as our example suggests. 

In a conceptually related mean-field study of spinless fermions on the honeycomb lattice with nearest-neighbor repulsive interaction of strength $V$ and in the Hofstadter regime -- i.e. adding a gauge field to produce fluxes through the plaquettes -- related effects were observed~\cite{shankar}. The non-interacting system has a nonzero Hall conductivity. Increasing $V$ and for two occupied bands, these bands were found to touch at a specific $V$ and a redistribution of Chern numbers led to a topological transition from Chern numbers (-1,1), in a topological ferrielectric phase, to Chern numbers (0,0) in a canonical ferrielectric phase. Other phase transitions involving changes in the Chern numbers increasing the repulsion $V$ can be found in Fig.~3 of Ref.\cite{shankar}.

Returning to our results, overall both dice and Lieb lattices behave very similarly, with the only exception that the lower three bands (out of the six bands of both models), namely the three bands that are populated at half filling, sometimes behave differently as a group. For the dice lattice, their combined Chern numbers add to zero in the entire phase diagram suggesting the absence of an Anomalous Quantum Hall Effect. However, for the Lieb lattice in weak coupling this does not occur and AQHE should be observable in  physical realization of the half-filled weakly-coupled Lieb lattices. In strong coupling, both Lieb and dice have the three lower bands cancelling their summed Chern numbers. Of course, merely by changing the chemical potentials, in both cases regions of nonzero AQHE can be easily found for both lattices.

The next computational challenge is the study of ribbons of dice and Lieb lattices employing the powerful DMRG technique fully incorporating quantum fluctuations, to confirm our results. 
At present, it is possible to study comfortably up to six legs in a ladder arrangement with DMRG. 
When we recently studied ribbons of the non-interacting dice lattice~\cite{soni20}, we observed that the physics of 
two-dimensional planes -- including Chern numbers deduced from the transverse Hall conductance $\sigma_{xy}$ -- 
can be rapidly reached by increasing the number of legs in ladders, at least for non-interacting electrons. 
$\sigma_{xy}$ can be calculated with DMRG as well. 
This study will be carried out in the near future.

\subsection*{Acknowledgments} We thank R.-X. Zhang for explaining us the second method to calculate Chern numbers in Lieb lattices, via only one unit cell. We thank N. Mohanta and C. Morais Smith for useful discussions.
All authors were supported by the U.S. Department of Energy (DOE), Office of Science, Basic Energy Sciences (BES), Materials Sciences and Engineering Division.

\section*{Appendix}

\subsection{Hartree vs Hartree-Fock Comparison}

\begin{figure}[!t] 
\centering
\includegraphics[width=3.1in, height=4.8in]{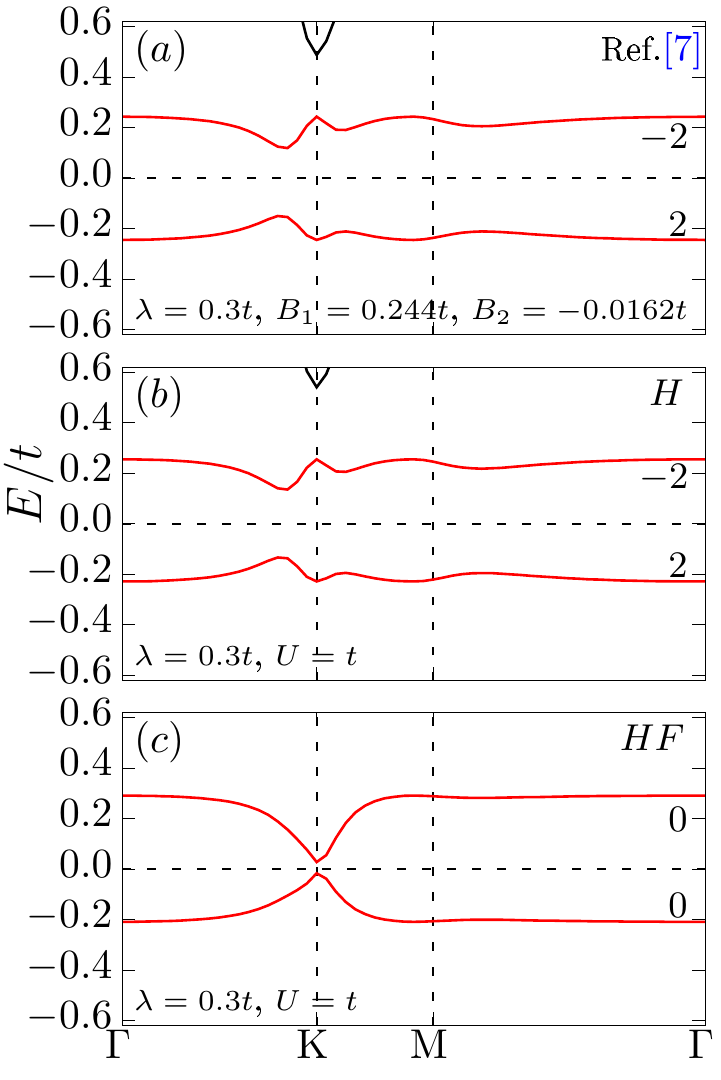}
\caption{Central energy bands near $E=0$ for the dice lattice at $U=t$, $\lambda = 0.3t$, and $\epsilon=0.6t$. $(a)$ depicts the energy bands reproduced from Ref.~\cite{wang11} for comparison, thus the label ``Ref.[7]''. For panel (a), those authors worked with the same couplings and evaluated the corresponding optimal magnetic fields for sublattices 1 and 2 via a variational method, with values shown as insets. The Chern numbers for the two red bands were $\pm2$. Panel $(b)$ illustrates the energy bands obtained by us when we only use the Hartree approximation. Here, the results are the same as in $(a)$, i.e. $C=\pm2$, and illustrates that using only Hartree is basically equivalent to optimizing external staggered fields, as intuitively expected. However, in $(c)$ we show the complete Hartree-Fock results of the present study. The energy in Hartree-Fock is lower than in Hartree. More importantly, note that the results change qualitatively, namely now the two red bands have $C=0$ and the gap is much smaller. \label{Fig: HF vs H Comparison}}
\end{figure}

In Fig. \ref{Fig: HF vs H Comparison} we compare results for the dice lattice using (a) non-interacting electrons in a staggered external field, (b) Hartree, and (c) Hartree-Fock methods at couplings $U=t$, $\lambda=0.3t$, and $\epsilon=0.6t$. The results for (a) are reproduced from Ref.~\cite{wang11} for the benefit of the readers. They were obtained optimizing variationally external magnetic fields associated with the two types of sites in the dice lattice, especifically
$B_1=0.224t$ and $B_2=-0.0162t$. In Fig.~\ref{Fig: HF vs H Comparison}$(b)$, we show our Hartree-only results for the same $U, \lambda, \epsilon$ parameters and realized that the Hartree results are quite similar to the variational results. The bands are almost identical. Moreover, in both cases the sum of Chern numbers for the first three bands from the bottom 
(only one shown) is $2$. 
However, in Fig. \ref{Fig: HF vs H Comparison}$(c)$ we show explicitly that when performing the full, and more accurate, HF approximation we obtain different results. Not only a lower ground-state energy is obtained in panel (c) than panels (a,b), but in addition the sum of Chern numbers for the first three bands is now $0$ in HF as opposed to 2 in just Hartree, showing that the Fock terms are relevant when $\lambda\neq0$. The main message is that the Fock terms are important in this context and they alter the physics qualitatively.

\FloatBarrier

\end{document}